\documentclass{article}

\usepackage{arxiv}

\usepackage[utf8]{inputenc} 
\usepackage[T1]{fontenc}    
\usepackage{hyperref}       
\usepackage{url}            
\usepackage{booktabs}       
\usepackage{amsfonts}       
\usepackage{nicefrac}       
\usepackage{microtype}      
\usepackage{xcolor}         
\usepackage{subfiles}
\usepackage{algorithm}
\usepackage{algorithmicx}
\usepackage[noend]{algpseudocode}
\usepackage{graphicx}
\usepackage{subfigure}
\usepackage{tabularx}
\usepackage{wrapfig}
\usepackage{amsmath,txfonts}
\usepackage{autobreak}
\newtheorem{assumption}{Assumption}
\newtheorem{theorem}{Theorem}
\newtheorem{lemma}{Lemma}

\newtheorem{proof}{Proof}
\newtheorem{definition}{Definition}
\title{DPDR: Gradient Decomposition and Reconstruction for 
Differentially Private Deep Learning}

\author{Yixuan Liu \\
        Renmin University of China\\
        \texttt{liuyixuan@ruc.edu.cn} \\
	\And
        Li Xiong \\
        Emory University\\
        \texttt{lxiong@emory.edu} \\
        \And
        Yuhan Liu \\
        Renmin University of China\\
        \texttt{liuyh2019@ruc.edu.cn} \\
        \AND
        Yujie Gu \\
        Kyushu University\\
        \texttt{gu@inf.kyushu-u.ac.jp} \\
        \And
        Ruixuan Liu \\
        Emory University\\
        \texttt{ruixuan.liu2@emory.edu} \\
        \And
        Hong Chen \\
        Renmin University of China\\
        \texttt{chong@ruc.edu.cn} \\
}

\begin{document}

\maketitle

\begin{abstract}
Differentially Private Stochastic Gradients Descent (DP-SGD) is a prominent paradigm for preserving  privacy in deep learning. It ensures privacy by perturbing gradients with random noise calibrated to their entire norm at each training step. However, this perturbation  suffers from a sub-optimal performance: it repeatedly wastes privacy budget on the general converging direction shared among gradients from different batches, which we refer as  common knowledge, yet yields little information gain. Motivated by this, we propose a differentially private training framework with early gradient decomposition and reconstruction (DPDR), which enables more efficient use of the privacy budget. In essence, it boosts model utility by focusing on incremental information protection and recycling the privatized common knowledge learned from previous gradients at early training steps. Concretely, DPDR incorporates three steps. First, it disentangles common knowledge and incremental information in current gradients by decomposing them based on previous noisy gradients. Second, most privacy budget is spent on protecting incremental information for higher information gain. Third, the model is updated with the gradient reconstructed from recycled common knowledge and noisy incremental information. Theoretical analysis and extensive experiments show that DPDR outperforms state-of-the-art baselines on both convergence rate and accuracy.
\end{abstract}

\section{Introduction}
Deep learning models achieve great success in various domains, but also pose privacy risks of the training data. For instance, adversaries are able to reconstruct original training data from model parameters \cite{fredrikson2015model, zhu2019deep}, and infer the membership of individuals in the training data from model outputs or gradients \cite{shokri2017membership,ye2022enhanced, jagielski2024students}. Differential Privacy (DP) \cite{dwork2006differential} is a standard privacy notion that introduces random noise to the computation, ensuring that the membership of any single data point remains undetectable from the output, thereby protecting individual privacy. To achieve DP for a deep learning model, Differentially Private Stochastic Gradient Descent (DP-SGD) \cite{abadi2016deep} is one of the most preeminent paradigms, which adds noises to gradients at each training step. The noise level scales up with the norm of entire gradients, which can significantly decrease model performance. To reduce noise amount, DP-SGD and recent variants typically bound the norm by clipping with adaptive threshold \cite{andrew2021differentially, pichapati2019adaclip} or scaling down gradients by normalization \cite{bu2024automatic}.

\begin{figure}
  \centering
  \label{pics_grad}
  \subfigure{
    \includegraphics[width=0.31\textwidth,trim=20 20 20 20,clip]{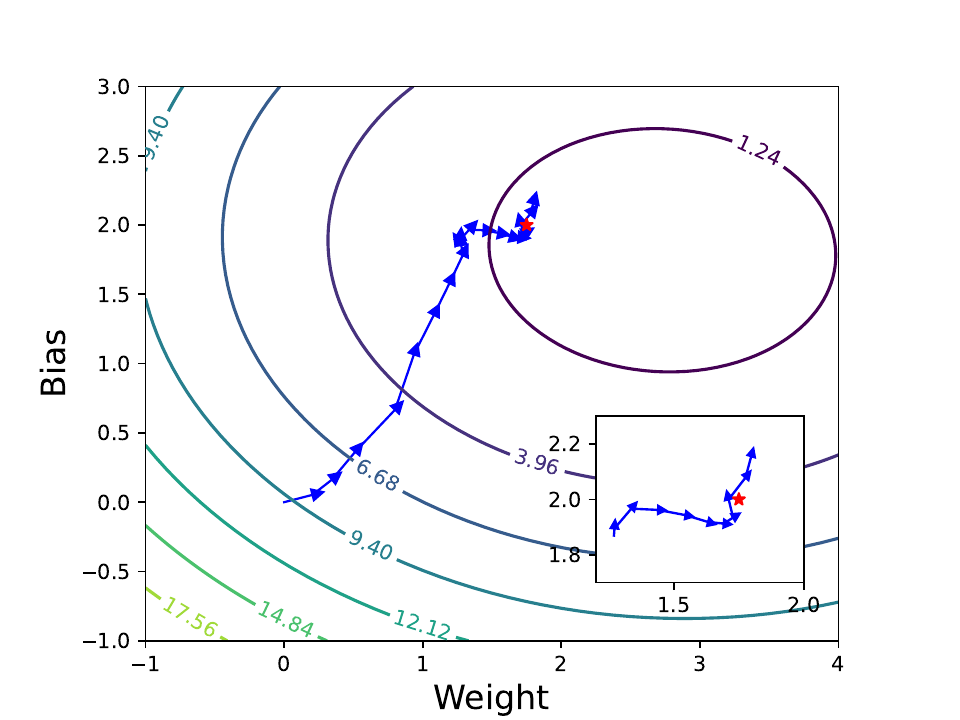}
    }
  \subfigure{
    \includegraphics[width=0.29\textwidth,trim=20 40 20 20,clip]{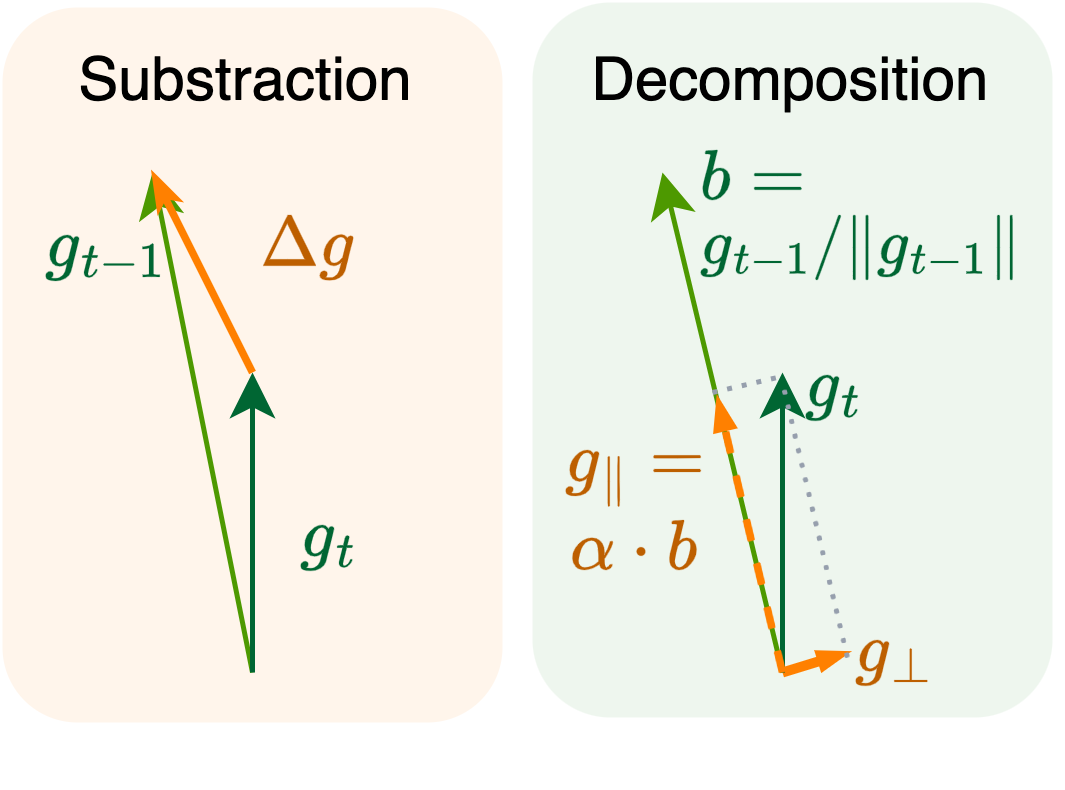}
    }
  \subfigure{
    \includegraphics[width=0.31\textwidth,trim=20 0 20 20,clip]{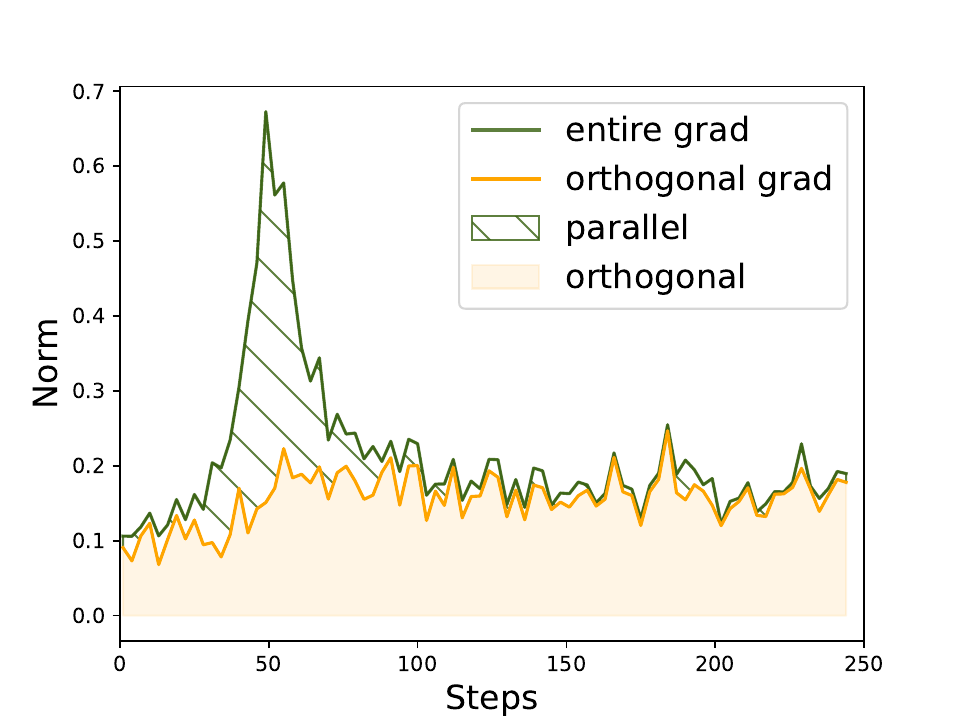}
    }
  \caption{\emph{Left}: SGD Visualization on linear regression model. Gradient directions are similar (coherent) at the early training stage, and fluctuate (stale) later. \emph{Middle}: In subtraction, incremental information is gradient difference $\Delta g=g_t-\tilde{g}_{t-1}$. In decomposition, incremental information is orthogonal gradient $g_\perp= g_t-\alpha\cdot b $, where $b$ is normalized ${g_{t-1}}$, parallel coefficient $\alpha=\langle g_t, b \rangle/\Vert b \Vert^2$. By Pythagorean Theorem, $\Delta g \leq g_\perp$. \emph{Right}: Norm of gradients on CIFAR10. At early stages, gradient norm fluctuates while orthogonal norm stays small and stable, which indicates the portion of common knowledge (green slash) is high compared to incremental information (orange range).}
\end{figure}

Improving privacy and utility tradeoff of DP-SGD is a well-recognized challenge. Existing works still suffer from a sub-optimal performance due to a common problem that a large amount of privacy budget is wasted on repeatedly protecting information that has already been learned from previous iterations. One of the key observations on gradients is that the gradients across different batches follow a similar direction especially in the early stages \cite{dai2018toward} (c.f. Fig. \ref{pics_grad}(Left)). The coherent direction could be regarded as the common knowledge shared by gradients over the whole dataset. Repeatedly collecting and protecting the common knowledge at different training steps leads to a large privacy budget consumption in return for little information gain.

Intuitively, by identifying the common knowledge from previous gradients and recycling it in the subsequent steps, we can significantly save privacy budget to only protect incremental gradient components complementary to the common knowledge for higher information gain.
A na\"ive solution is subtracting the previous noisy gradient from the current one and perturbing only the difference (c.f. Sec. \ref{sec_grad}). However, the difference may not remove all common knowledge and may suffer from a norm even larger than the original gradient norm, leading to more injected noises. Therefore, characterizing the common knowledge precisely to keep the bounded norm of incremental information as small as possible is a challenging problem.

\begin{figure}
  \centering
  \label{pics_framework}
  \includegraphics[width=0.9\linewidth,trim=30 20 30 20,clip]{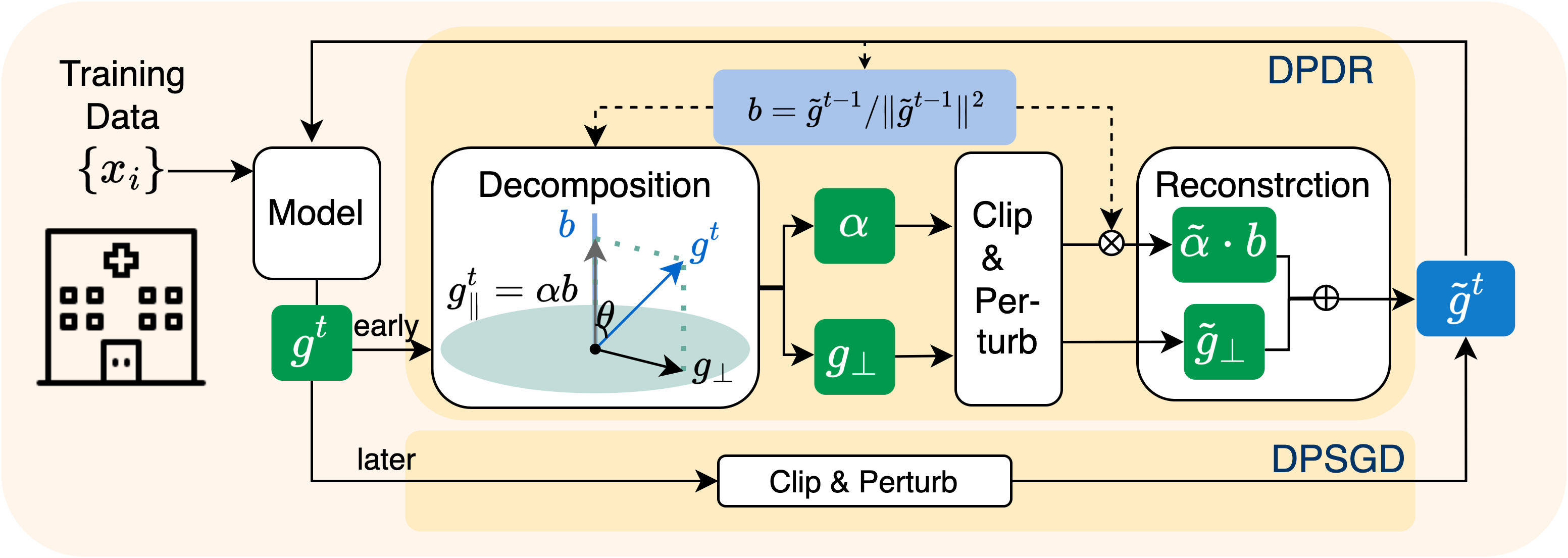}
  \caption{Framework of DPDR. First, it decompose current gradient $g_t$ into $g_\perp$ (incremental information) and $\alpha \cdot b$ by directional decomposition based on previous normalized noisy gradients $b$ (common knowledge). The parallel coefficient $\alpha$ and $g_\perp$ are perturbed for further reconstruction with $b$. Model is updated based on reconstructed gradient}
\end{figure}

To this end, we propose \textbf{DPDR}, a \underline{\textbf{D}}ifferentially \underline{\textbf{P}}rivate training framework with gradient \underline{\textbf{D}}ecomposition and \underline{\textbf{R}}econstruction at early stage as shown in Fig. \ref{pics_framework}.
Specifically, it consists of a private Gradient Decomposition and Reconstruction technique (GDR) and a mixed strategy. For GDR, it first directionally decomposes gradients into two parts: orthogonal components $g_\perp$ and parallel components $g_\varparallel$ based on noisy previous gradients (c.f. Fig. \ref{pics_grad} (Middle)). The extracted incremental information $g_\perp$ is completely independent of common knowledge and achieves a smaller norm due to Pythagorean Theorem. Then most privacy budget is spent on perturbing $g_\perp$ with bounded norm, and only a small privacy budget is used on parallel coefficient $\alpha$ for recycling common knowledge. 
At last, we recover the whole gradients by summing up noisy incremental information and common knowledge, which ensures a correct model converging direction and accelerates the convergence rate.


Furthermore, the mixed strategy applies GDR at the early training steps and switches to DP-SGD later. As the large proportion of common knowledge that GDR benefits from mainly appears at early stages (c.f. Fig. \ref{pics_grad} (Right)), it is unnecessary to spend privacy budget on recycling common knowledge when it is too little at later stages. Switching to DP-SGD allows full use of privacy budget.



\vspace{-0.2cm}

Our main contributions are summarized as follows:
\begin{itemize}
	\item We develop a directional-decomposition-based privatization technique for DP-SGD.  It provides a higher information gain with less noise injection by (1) spending most of the privacy budget on the incremental information in current gradients, and (2) reusing  the common knowledge ~(a general converging direction) from historical gradients. 
        \item We design a mixed training framework DPDR based on a universal pattern that gradients from the early training steps are more alike to each other. It promises a better performance by making the most of the privacy budget for obtaining more informative knowledge.
	\item We theoretically prove that compared to DP-SGD, our proposed methods promise a faster convergence rate benefiting from the reusing of common knowledge and less noise injection under the same level of privacy guarantee.
	\item Our extensive experiments on real-world datasets confirm that DPDR outperforms DP-SGD and its SOTA variants on both convergence rate and model accuracy.

\end{itemize}

\section{Related Work}
DP-SGD proposed in \cite{abadi2016deep} develops as a predominant differential private model training framework in deep learning. Many works have made efforts to improve the utility from different angles. Different from these works, we focus on the fundamental operation, the perturbation on common knowledge, which is rarely noticed. As a result, our method could be regarded as a building block for most advanced DP-SGD variants.

\textbf{Clipping strategy in DP-SGD.} Clipping is introduced in DP-SGD to bound the sensitivity of gradients. A larger clipping bound brings to large noise amount, while smaller bound leads to bias. Therefore, the effects of clipping is analyzed recently for formal trade-off between them \cite{chen2020understanding, xiao2023theory}. To reduce noise amount, several works tunes clip bound by data-dependent strategy\cite{andrew2021differentially} , or normalize gradients for smaller gradient norm \cite{bu2024automatic, yang2022normalized}. These work improve DP-SGD with a better noise scale, while keep the relationship between noise level and whole gradient norm. 

\textbf{Adaptive Optimization} Advanced optimizers such as Adam, Neterov boost the effectiveness of SGD \cite{nesterov2013gradient,kingma2014adam}. However, differential private noises comprise the performance of these optimizers seriously, as noises accumulate in the preconditioner along iterations. Averaging and adaptive strategy \cite{lin2022dynamic, zhou2020private,xiao2023theory, li2022differentially} are applied on preconditioner to decrease the variance of historical noises. While they sometimes still demonstrate comparable performance with the original DP-advanced-optimizer \cite{tang2024dp}.

\textbf{Projection and Heuristic Improvements} Recent works attempt to project gradients into certain space to avoid high dimension curse \cite{zhou2020bypassing}, or predefined space \cite{asi2021private}. An assumption is made on gradient distribution with auxiliary information or prior knowledge. Another line of works made improvements with better gradients selection \cite{fu2023dpsur}. These works are orthogonal with our method as none of them consider the internal noise design in one gradient. We notice that a recent work \cite{murata2023diff2} proposes adding noises to the difference of consecutive sanitize gradients on the same batch, which is similar with our strawman approach. It performs well on low dimension datasets, while the calculation cost doubles for gradient recomputing, and the bias of reconstructing gradients is not considered.

\section{Preliminaries}
In this section, we recap the notions related to differential privacy and the framework of DP-SGD. First, we introduce the privacy notion, Differential privacy (DP) \cite{dwork2006differential}, a de facto standard that is widely accepted to provide rigorous privacy for raw data. The formal definition is as follows.
\begin{definition}[Differential Privacy]
	For any $\epsilon, \delta \geq 0$, a randomized algorithm $M: \mathcal{D} \rightarrow \mathcal{R}$ is $(\epsilon, \delta)$-differential privacy if for any neighboring datasets $D, D' \in \mathcal{D}$ and any subsets $S \subseteq \mathcal{R}$,
	$$\Pr[M(D) \in S] \leq e^{\epsilon}\Pr[M(D') \in S] + \delta.$$
\end{definition}
The DP guarantee for the function $f: \mathcal{X}^n \rightarrow \mathbb{R}^d $ is implemented by adding random noises. The noise scale is determined by privacy budget $\epsilon$ and sensitivity $\Delta f$.

\begin{definition}[Sensitivity]
The $l_s$-sensitivity 
of a function $f: \mathcal{X}^n \rightarrow \mathbb{R}^d $ is
$\Delta f = \mathop{\max}\limits_{x,x'\in \mathcal{X}^n} \Vert f(x) - f(x') \Vert_s.$
\end{definition}
Sensitivity captures the worst-case changes of outputs when a single input sample differs. In deep learning, we usually adopt $l_2$ norm as metric. Additionally, following property allows us ensure privacy of arbitrary post operation on perturbed sensitive data.
\begin{lemma}[Post-processing]
Let $M: \mathcal{X}^n \rightarrow \mathcal{R} $ be a randomized algorithm that satisfies $(\epsilon, \delta)$-DP, $f: \mathcal{R} \rightarrow \mathbb{R'} $ be an arbitrary function. Then $f \circ M : \mathcal{X}^n \rightarrow \mathbb{R'}$ is also  $(\epsilon, \delta)$-DP.
\end{lemma}

DP-SGD provides a general scheme for private deep learning. Concretely, a small batch of sample $L_t$ is randomly selected from the whole datasets $D$ with probability $\frac{B}{|D|}$, where $B$ is batch size. To protect the averaged gradient of a batch at each training step, DP-SGD clips per-sample gradient with pre-defined clipping bound $C$, then adds noises to the sum of clipped gradients with sensitivity $C$:  
$$ \tilde{g}_t = \frac{1}{B} (\sum_{x_i \in L_t} g_t(x_i) / \max(1, \frac{\Vert g_t(x_i)\Vert_2}{C}) + N(0, \sigma^2 C^2 \boldsymbol{I})) $$
Where $\sigma$ is the noise scale determined by privacy budget $\epsilon$. By clipping and perturbing, model release at each step is protected. Composition theorem is used for accounting privacy consumption during $T$ epochs, RDP\cite{mironov2017renyi} and Moment Accountant\cite{abadi2016deep} are usually adopted.  

\section{Proposed Methods}
In this section, we demonstrate the proposed framework DPDR. 
We first show the observation on common knowledge brought by coherent gradients across steps at the early SGD training process, and introduce a strawman approach to recycle it for less privacy budget waste in Section \ref{sec_grad}. Then we introduce a decomposition and reconstruction technique to completely disentangle the common knowledge from noisy gradients, with a mixed strategy involving DP-SGD for more effective use of privacy budget in Section \ref{sec_dpedr}.

\subsection{Gradient Variation in Vanilla SGD}
\label{sec_grad}
\begin{wrapfigure}{r}{4cm}
\label{pics_grad_diff}
\centering
\includegraphics[width=0.28\textwidth, trim=30 20 30 30,clip]{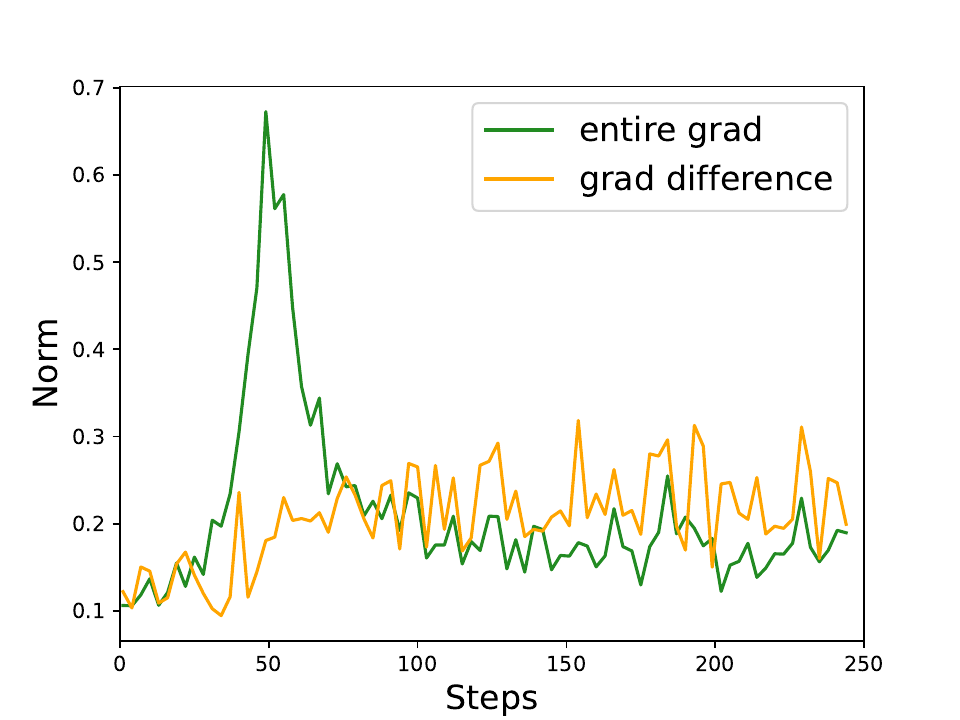}
\caption{The norm of difference may be larger than the original gradient. CIFAR10.}
\end{wrapfigure}

DPDR boosts utility by taking advantage of characteristics of gradients which are inherently similar to each other at certain training stage. Specifically, the gradients of stochastic batches of samples are computed at each training step in SGD, which show two characteristics: coherence and staleness. Coherence means that gradients overall maintain similar across consecutive steps to a certain direction due to the similarity of samples, which is also observed by some works \cite{dai2018toward, chatterjee2020coherent}. On the other hand, staleness captures the difference of gradients across steps, which usually appear at two steps far from each other or at stages when gradient directions change rapidly. \cite{li2022differentially}. 

A key observation in our work is that gradients are coherent at the early stage of the training process, while easier to stale at the later stage. As shown in Fig. \ref{pics_grad} (Left), gradients first follow similar directions. The general direction indicates that common knowledge repeatedly appears in each gradient at early training stage. At later stage, fluctuating gradient direction suggests less common knowledge preserved.
Thus, reusing the common knowledge shared among recent steps allows us to avoid repeatedly collecting and perturbing on general direction learned already, and save privacy budget to protect the incremental components which is the more informative part in current gradients. 

A strawman solution is subtracting the previous noisy direction from current gradients, then adding noises to the difference and recovering it by adding the previous noisy direction as Eq.\eqref{eq_diff}.
\begin{equation}
\label{eq_diff}
  \tilde{g}_t =  \frac{1}{B}\sum_{i=1}^B Clip((g_t(x_i)-\tilde{g}_{t-1}), C) + N(0,C^2\sigma^2) +  \tilde{g}_{t-1}  
\end{equation}
On the surface, this solution filters out the common knowledge  ($\tilde{g_{t-1}}$) from current gradients and exploits all privacy budgets to protect only the incremental information (difference of gradients). However, it may not remove all common knowledge completely from current gradient. For example, cosine similarity between the $\tilde{g}_{t-1}$ and difference probably is nonzero in Fig\ref{pics_grad} (Middle). Therefore, when the staleness between two consecutive gradients grows up, the difference suffers from a large norm than expected ~(c.f.Fig. \ref{pics_grad_diff}), leading to unnecessary noise injecting.

\subsection{Directional Decomposition and Reconstruction}
\label{sec_dpedr}
\begin{wrapfigure}{r}{4cm}
\label{pics_grad_perp_dist}
\centering
\includegraphics[width=0.28\textwidth, trim=30 20 30 37,clip]{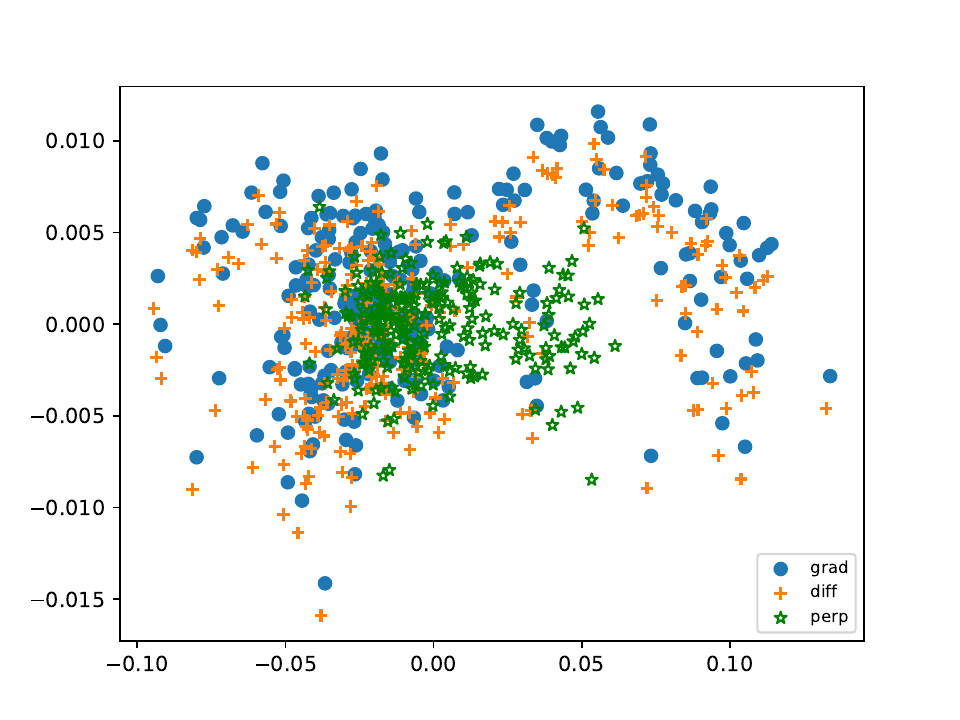}
\caption{Distribution of orthogonal components of gradient are more concentrated compared to difference. CIFAR10.}
\end{wrapfigure}
\vspace{-0.3cm}

In this section, we state the whole construction of DPDR~(c.f. Fig. \ref{pics_framework}.), which consists two parts: (1) the Gradient directional Gecomposition and Reconstruction technique (GDR) and (2) the mixed strategy, where the decomposition technique is applied only at the early training steps for a higher information gain.

\textbf{Gradient Decomposition and Reconstruction Technique} 
GDR decomposes gradients based on the previous gradients and extracts an orthogonal component completely independent with common knowledge. According to the property of vector decomposition, the orthogonal gradient is smaller than full gradient (c.f. Fig. \ref{pics_grad_perp_dist}), hence less noises is injected with smaller sensitivity. To maintain the full information from original samples, GDR reconstructs noisy gradients by recycling previous noisy gradients. 

As shown in Algorithm \ref{algo}, immediately after the first step. Gradient $g_{t}(x_i)$ is decomposed into orthogonal components ${g_{t}(x_i)}_\perp$ and parallel component $\alpha_t(x_i)\cdot b$ based on the common knowledge vector $b$ computed from the previous steps. The parallel coefficient $\alpha_t(x_i)$ quantifies the amount of common knowledge (c.f. Eq.\eqref{eq_decompose}). The whole technique is applied layer by layer. A key operation is formalized as follows:
\begin{equation}
\label{eq_decompose}
    \alpha_t(x_i) = \langle g_{t}(x_i), b\rangle / \Vert b \Vert^2_2 \text{,}\quad g_{t}(x_i)_\perp = g_{t}(x_i) - b \cdot \alpha_t(x_i)\text{, }\quad\text{where } b=\tilde{g}_{t-1}/\Vert \tilde{g}_{t-1} \Vert
\end{equation}

We then guarantee differential privacy for Algorithm \ref{algo}. At the first step, the entire gradient is clipped by $C_g$ and perturbed. At the following steps, we protect both $g_{t}(x_i)_\perp$ and $\alpha_t(x_i)$ by clipping and perturbing separately.  
The perturbed gradient components are reconstructed by adding $b \cdot \tilde{\alpha}_t(x_i)$ back for the following training. At last, $b$ is also updated with a normalized reconstructed gradient.

\textbf{Mixed Strategy} Furthermore, we propose a mixed strategy by applying the GDR at the early training steps and maintain DP-SGD later. 
By spending a small portion of privacy budget on parallel coefficient $\alpha$, GDR benefits from the reusing common knowledge with little price, and focuses on incremental information protection. The model gain drops when the proportion of common knowledge in current gradients decreases. As mentioned, gradients are coherent early and stale later, hence the proportion of common knowledge is larger early and goes down later (Fig. \ref{pics_grad} (Right)). Hence at later training stage when the proportion approaches zero, protecting $\alpha$ and reusing common knowledge wastes privacy budgets. As a result, we switch to DP-SGD for better utility and higher efficiency. 

\textbf{Discussion} The proposed method, DPDR, is regarded as a fundamental building block for private deep learning, which achieves smaller noise levels when gradients are more coherent and less stale at consecutive steps. (1) It is noticed that $\alpha$ retains magnitude information of $g_{t-1}$, which measures the amount of common knowledge that needs to recycle. As $\alpha$ only holds constant-level dimensions, the noise amount injected to it is irrelevant to model parameter dimensions, thereby has little affects on utility. Hence a large portion of privacy budget is assigned to orthogonal components ${g_{t}(x_i)}_\perp$ for higher information gain and better performance. (2) Early stage is a range of training steps. Though the range varies when the dataset and model change, the model performance is not sensitive to the number of step under a large range.

\algnewcommand{\LineComment}[1]{\State \(\triangleright\) #1}
\begin{algorithm}[tb]
	\caption{Decomposition and Reconstruction: $DPDR(\cdot)$}
	\label{algo}
	\begin{algorithmic}
	\Require $T$, $s$, $\{x_i\}_{i \in [n]}$, $L(w)$, $C_\alpha$,$C_\perp$, $C_g$, $\sigma_\alpha$,$\sigma_\perp$,$\sigma_g$, learning rate $\gamma$, batch size $B$.
		\Ensure model $w$
		\State Model initializes $w_0$, $b = w_0/\Vert w_0 \Vert$.
			\For{$t\in [T]$} 
                    \If{$t\in [2,...,s]$}   \Comment{Decomposition after first step}
				\State Random sample $\{x_i\}_{i \in L_{t}}$ with sampling ratio $B/|D|$
				\State For each $i \in L_t$, compute $g_{t}(x_i) \leftarrow \nabla L(w_{t}, x_i)$
				\LineComment{\texttt{\textbf{Directional Decomposition}}}
                \State{\textcolor{black}{$g_{t} (x_i)_\perp, \alpha_{t}(x_i) \leftarrow \Pi_{b} (g_{t} (x_i))$   \Comment{Projection base: noisy grad}}}
				\LineComment{\texttt{\textbf{Clip \& Perturbation}}}
				\State $\tilde{g}_{t \perp}  \leftarrow \frac{1}{L}(\sum_i \frac{g_{t} (x_i)_\perp}{\max(1, \lVert g_{t} (x_i)_\perp \rVert / C_\perp)} + \mathcal{N}(0, \sigma_\perp^2 C_{\perp}^2 \boldsymbol{I}))$ 
				\State $\tilde{\alpha}_{t}  \leftarrow \frac{1}{L}(\sum_i \frac{\alpha_{t}(x_i)}{\max(1, \alpha_{t}(x_i)/C_2)} + \mathcal{N}(0, \sigma_\alpha^2 C_\alpha^2))$ \Comment{Parallel factor}
                    \LineComment{\texttt{\textbf{Directional Reconstruction}}}
				\State $\tilde{g}_{t} \leftarrow \tilde{\alpha}_{ t} \cdot b + \tilde{g}_{t\perp} $ 
				\State $w_{t+1} \leftarrow w_{t} - \gamma \tilde{g}_{t}$
                \LineComment{\texttt{\textbf{Update Parallel Base}}}
                \State{\textcolor{black}{$b \leftarrow \tilde{g}_{t}/\Vert\tilde{g}_{t}\Vert$}}  \Comment{\textcolor{black}{Noisy Base}}
                \Else
                    \State DP-SGD$(w_t, \sigma_g, C)$ \Comment{Alternative when gradient coherence reduce}
                 \EndIf
			\EndFor
		\Return $w_T$
	\end{algorithmic}
\end{algorithm}

\section{Privacy Analysis}
In this section, we first demonstrate the privacy guarantee for Algorithm \ref{algo}, and explain the reason that noises scale introduced in DPDR is smaller than DP-SGD. 

In Algorithm \ref{algo}, the entire gradient is accessed and protected at the first step. At the following steps, both $g_\perp$ and $\alpha$ are clipped and perturbed separately. 
After early decomposition, DP-SGD is adopted at later stages, which is also differential private. 
\begin{theorem}[Privacy Guarantee of Algorithm \ref{algo}]
\label{th_priv}
    There exists constants $v_1$,$v_2$,$v_3$,  batch size $B$, dataset size $|D|$, clipping bound $C_\perp$, $C_\alpha$, $C_g$ and training steps $T$ such that for any $\delta>0$, $\epsilon_\perp < v_1B^2/|D|^2T$, $\epsilon_\alpha < v_2B^2/|D|^2T$, $\epsilon_\perp+\epsilon_\alpha < v_3 B^2/|D|^2T$, if noise multipliers satisfy $\sigma_\perp^2 \geq \frac{v_1|B|^2T\ln(1/\delta)}{N^2\epsilon_\perp^2}$, $\sigma_\alpha^2 \geq \frac{v_2^2|B|^2T\ln(1/\delta)}{N^2\epsilon_\alpha^2}$ and $\sigma_g^2 \geq \frac{v_3^2|B|^2T\ln(1/\delta)}{N^2(\epsilon_\alpha+\epsilon_\perp)^2}$, Algorithm 1 is $(\epsilon_\perp+\epsilon_\alpha, \delta)$-DP.
\end{theorem}

The noise scale of DPDR is much smaller compared to DP-SGD. Though noise scale depends on the clipping bound, smaller sensitivity allows lower clipping bound with same clipping bias. In this section, we prove that the sensitivity is smaller. The effects of clipping bound selection will be discussed in next section.
Specifically, sensitivity of gradients in DPDR contains two parts:
\begin{align}
    \nabla L(w_t)= {\nabla L(w_t)}_\perp + {\nabla L(w_t)}_\parallel 
    = \underbrace{{\nabla L(w_t)} - \Pi_{\tilde{\nabla L}(x_{k-1})} (\nabla L(w_t))}_{\Delta f_\perp}  + \underbrace{\Pi_{\tilde{\nabla L}(x_{k-1})} (\nabla L(w_t))}_{\Delta f_\alpha} 
\end{align}
Under the assumption that gradient satisfies $\rho$-smoothness (c.f. Assumption \ref{assumption_smooth}), based on triangle properties and Chernoff inequality, we achieve upper bound of sensitivity separately. The full proof are presented in Appendix \ref{appendix_sensitivity}.
\begin{equation}
\label{eq_sensitivity_perp}
	\Delta f_\perp 
	\leq \min( \Vert \nabla L(w_t) \Vert,  2\rho \Vert w_t - w_{t-1} \Vert)
\end{equation}
\begin{equation}
    \Delta f_\alpha \leq \Vert \cos \theta \cdot \nabla L(x_{t}) \Vert
    \leq \Vert \nabla L(w_{t}) \Vert
\end{equation}
For $\Delta f_\perp$, it is noticed that the setting in this work with real-world datasets is bounded by $2\rho \Vert w_t - w_{t-1}$ which is far less than $\Vert \nabla L(w_t) \Vert$ with at least $99\%$ probability. Thus, a smaller clipping bound on $g_\perp$ is allowed at very low price of clipping bias compared to DP-SGD. 

For $\Delta f_\alpha$, the sensitivity is no more than entire gradient norm. Though $\Delta f_\alpha$ is not as smaller as $\Delta f_\perp$, it makes far less affects on performance as it is not the dominant term on noise variance. As noises on $\alpha$ scale up with the number of model layers $m$ rather than dimension $d$, where $m \ll d$.

\section{Convergence Analysis}
In this section, we provide convergence analysis of proposed method DPEDR for non-convex smooth optimization. The effects of noises introduced by DP guarantee is analyzed, the per-sample clipping strategy on both orthogonal components and parallel coefficient $\alpha$ is considered as well.

\begin{assumption}[$\rho$-Smoothness]
\label{assumption_smooth}
The loss function is $\rho$-smooth. for any $w, w' \in \mathbb{R}^d$ and batch samples $x=[x_1, x_2, ..., x_B]$, we have $\Vert \nabla \mathcal{L}(w,x) - \nabla \mathcal{L}(w',x) \Vert \leq \rho\Vert w-w' \Vert$.
\end{assumption}


\begin{lemma}[Convergence without clipping bias] If the orthogonal components $g_\perp$ and parallel coefficient $\alpha$ are clipped by $C_\perp$ and $C_\alpha$, sampling ratio as $q=B/|D|$, learning rate as $\gamma$. over the $T$ iteration, DPDR ensures that for $t=1,2,...,T$,
\label{lemma_dpedr_noclip}
\begin{small}
\begin{align*}

&\mathbb{E}[\Vert \nabla\mathcal{L}(x_{t-1}) \Vert^2] 
 \leq \frac{1}{\gamma T}\mathcal{L}(w_{0}) + \mathcal{O}(\rho\gamma d C^2_\perp \sigma^2_\perp ) \vphantom{\frac{1}{\gamma T}}.


\end{align*}
\end{small}
\end{lemma}
The utility loss of DP-SGD described in Lemma \ref{lemma_dpedr_noclip}, which is dominated by perturbation on orthogonal components $g_\perp$. Clipping bound $C_\perp$ is crucial, which directly enlarge noise scale. While we cannot choose as small clipping bound as we can, since the bias is introduced into gradients due to clipping. To analyze the trade-off between DP noise and clipping, we provide the utility loss for DPDR below.


Then we formalize gradient operation as 
$ \tilde{g}_t =  \frac{1}{B} (\sum^B_{i=1}Clip(\alpha_t, C_\alpha)+N(0, C^2_\alpha\sigma^2_\alpha)) \cdot b + \frac{1}{B}(\sum^B_{i=1}Clip(g_{t\perp}, C_\perp)+N(0,C^2_\perp\sigma^2_\perp))$. 
Next we make assumption on sampling noises caused by stochastic gradient distribution. Along with decomposition, the sampling noises on gradients are decomposed into the same directions. Hence we have $\xi_t=\xi_{\perp,t}+\xi_{\varparallel,t}$, and $\Vert \xi_t \Vert^2=\Vert \xi_{\perp,t}\Vert^2+\Vert \xi_{\varparallel,t}\Vert^2$. A minimal assumption on sampling noises  is defined in Assumption \ref{assumption_grad_moment}, which is a generally adopted by recent works \cite{xiao2023theory,chen2020understanding}. 

\begin{assumption}[Bounded Second Moment of Stochastic Gradient]
\label{assumption_grad_moment}
For and given dataset $D=\{x_1, x_2, ..., x_n\}$, loss function $L(w)=\frac{1}{n}\sum_{i=1}^n l(w, x_i)$ for a random record $x_i$ sampling from $D$, the sampling noise is bounded by $\tau^2$, after decomposition ,the sampling noises are bounded by $\tau_\perp^2$ and $\tau_\alpha^2$, i.e., $ \mathbb{E}_{x_i \in D} [\Vert \nabla L(w, x)-\nabla l(w,x) \Vert^2] \leq \tau^2 $, $ \mathbb{E}_{x_i \in D} [\Vert \nabla L_\perp(w, x)-\nabla l_\perp(w,x) \Vert^2] \leq \tau_\perp^2 $, $ \mathbb{E}_{x_i \in D} [\Vert \nabla L_\alpha(w, x)-\nabla l_\alpha(w,x) \Vert^2] \leq \tau_\alpha^2 $.
\end{assumption}

\begin{theorem}[Convergence with clipping threshold] 
\label{theorem_DPEDR}
Set clipping bound on $\alpha$ as $C_\alpha$ and orthogonal components as $C_\perp$, and the probability of clipping as $P_{\perp,t} = \Pr[\xi_{\perp,t}\in S_{\Vert \nabla\mathcal{L}_\perp(x_{t-1}) +\xi_{\perp,t}\Vert \geq C_\perp}]$, as $P_{\varparallel,t} = \Pr[\xi_{\varparallel,t}\in S_{\Vert \nabla\mathcal{L}_\perp(x_{t-1}) +\xi_{\varparallel,t}\Vert \geq C_\alpha}]$ separately, sampling ratio as $q=B/|D|$, learning rate as $\gamma$, $\gamma'=B\gamma$. over the $T$ iteration, DPEDR ensures that for $t=1,2,...,T$,
\begin{small}
\begin{align*}

&\mathbb{E}[(1-P_{\perp, t})\Vert \nabla\mathcal{L}(x_{t-1}) \Vert^2   
+ \Vert \nabla\mathcal{L}_\perp(x_{t-1}) \Vert (\frac{C_\perp P_{\perp,t} }{4}- \sqrt{P_{\perp,t}} \tau_{\perp,t})+ \Vert \nabla\mathcal{L}_\varparallel(x_{t-1}) \Vert \cdot(\frac{C_\varparallel P_{\varparallel,t} }{4} - \sqrt{P_{\varparallel,t}} \tau_{\varparallel,t})] \\
& \leq \frac{1}{\gamma T}\mathcal{L}(w_{0}) + \frac{\rho\gamma'}{2}(2C_\perp^2+dC_\perp^2\sigma_\perp^2+2C_\alpha^2+mC_\alpha^2\sigma_\alpha^2) 
+\frac{15}{4T}\sum_{t=1}^T\mathbb{E}[C_\perp\tau_{\perp,t}\sqrt{P_{\perp,t}} + C_\alpha\tau_{\alpha,t}\sqrt{ P_{\alpha,t} }] \\

& \leq \underbrace{\frac{1}{\gamma T}\mathcal{L}(w_{0})}_{\text{general term of SGD}} + \underbrace{\mathcal{O}(\rho\gamma d C^2_\perp \sigma^2_\perp) \vphantom{\frac{1}{\gamma T}}}_{\text{by DP noises}} 
+ \underbrace{\frac{15}{4T}\sum_{t=1}^T\mathbb{E}[C_\perp\tau_{\perp,t}\sqrt{P_{\perp,t}} + C_\alpha\tau_{\alpha,t}\sqrt{ P_{\alpha,t} }]}_{\text{by clipping}}

\end{align*}
\end{small}

\end{theorem}
The utility loss of DPDR with clipping bias is provided in Theorem \ref{theorem_DPEDR}. The second term \emph{DP noises} mainly caused by the perturbation on gradients after decomposition. Though both directions are perturbed, notice that the number of model layers $m$ is far less than the number of model parameters $d$, the noises is dominated $\mathcal{O}(dC_\perp^2\sigma_\perp^2)$ on orthogonal direction. The last term \emph{clipping} demonstrates the bias due to clip on both directions characterized by $\mathcal{O}(C_\perp + C_\alpha)$.

According to Lemma \ref{lemma_dpedr_noclip}, a small bound reduces the noises. However, Theorem \ref{theorem_DPEDR} indicates that small bound severely slows down the convergence as clipping bias increases with larger probability of clipping bias $P_{\perp,t}$. Compared with DP-SGD, DPDR alleviates such degradation by decomposition. (1) As mentioned, the norm of dominant variable $g_\perp$ is smaller than $g$. As a result, choosing clipping bound the same as DP-SGD leads to same noise scale, while decreases the probability of clipping bias, thereby achieves faster convergence rate in practice. In other word, at the same bias level, DPDR is allowed to select smaller $C_\perp$ for less noises. (2) Clipping bound $C_\alpha$ is not in dominant term, hence larger $C_\alpha$ is allowed with low clipping bias probability.

\section{Experiment Results}
In this section, we demonstrate the experiment results to verify the accuracy enhancement and convergence rate of the proposed method DPDR on public datasets and classic deep learning models. 

\textbf{Datasets and Models} We conduct experiments on datasets MNIST\cite{lecun1998mnist}, CIFAR-10\cite{krizhevsky2009learning} and SVHN\cite{netzer2011reading} with model 4-layer CNN, 5-layer CNN, ResNet18 separately, by group normalization and cross-entropy loss function for all of them. The noise scale is derived under privacy budget $\epsilon=3$ and $8$ with fixed $\delta=10^{-5}$. All of the results are presented with the best results after the parameter tuning. As the hyperparameter tuning process is a common practice for all private machine learning methods \cite{koskela2024practical,papernot2021hyperparameter}, we don't account for the privacy loss in this paper. The setting details are presented in Appendix \ref{appendix_exp}.

\textbf{Baseline} We compare the DRDP with DP-SGD\cite{abadi2016deep} and its state-of-the-art variants improved on clipping or adding noises: AutoClip\cite{bu2024automatic}, DIFF2\cite{murata2023diff2}, DPAdam, DIFF. AutoClip removes the influence of clipping threshold by normalization, which represents the line of clipping enhanced work. As mentioned in \cite{tang2024dp}, DPAdam performs even better on CIFAR-10 and similar datasets than DPAdam-variants, hence we adopt DPAdam as an important baseline. Additionally, we also demonstrate the performance of the strawman approach introduced in Section \ref{sec_grad}, and name it as DIFF.


\subsection{Overall Performance Evaluation}
Tab. \ref{tab_acc} demonstrates the overall performance of DPDR and baselines. (1)The proposed method DPDR achieves higher accuracy compared to existing works across all datasets and privacy budgets. We notice that on larger datasets CIFAR-10 and SVHN the improvement are more significant, around 2\% higher than baselines. The less promotion on MNIST is reasonable as it has almost approached non-private accuracy. The enhancement of DPDR benefits from less noise introduced in training process with smaller norm at the early training stage, and maintain precision at the following phases. 
(2) Consistent with our explanation above, DPAdam and Autoclip achieve similar accuracy with DP-SGD as they perturb gradient with noises calibrated to the norm of entire gradients, the ratio between information and noises injecting did not change. (3) An upward trend in accuracy is seen on DIFF2 and DIFF as the model size decreases. As a smaller model usually enjoys more coherent gradients, leading to a much smaller  norm.
On the contrary, DPDR is more robust to gradient variation, hence performs stable on different model sizes.

\begin{table}
  \caption{Accuracy Comparison on public datasets after paramter tuning.}
  \label{tab_acc}
  \centering
  \begin{tabular}{*{5}{p{0.15\textwidth}<{\centering}}}
    \toprule
     $\epsilon$   & Method & MNIST & CIFAR-10 & SVHN \\
    \midrule
    $3$ &\textbf{DPDR} & \textbf{96.42\%} & \textbf{57.14\%} &  \textbf{67.44\%}  \\
     & AutoClip  & 96.15\% & 55.25\%  & 65.87\%  \\
     & DIFF      & 95.65\% & 52.02\% & 13.60\%   \\
     & DIFF2  & 95.91\% & 55.86\%  & 65.84\% \\
     & DPAdam    & 96.31\% & 55.30\%  & 65.61\%  \\
     & DP-SGD  & 96.16\% & 55.48\% &65.87\%   \\
    \midrule           
    $8$  &\textbf{DPDR} & \textbf{97.03\%} & \textbf{61.99\%} & \textbf{72.12\%}   \\
     	& AutoClip  & 96.85\% & 59.81\%  & 72.03\%  \\   
    	& DIFF      & 96.54\% & 52.58\% & 13.79\%   \\
            & DIFF2     & 96.05\% & 60.02\% & 69.07\%   \\
            & DPAdam    & 96.53\% & 58.69\% & 71.15\%   \\
            & DP-SGD     & 96.62\% & 59.85\% & 72.05\%   \\
    \bottomrule
  \end{tabular}
\end{table}
\vspace{-0.2cm}

\subsection{Convergence Evaluation}
\begin{figure}
  \centering
  \label{pics_conv}
  \includegraphics[width=0.32\textwidth, trim=10 5 30 10,clip]{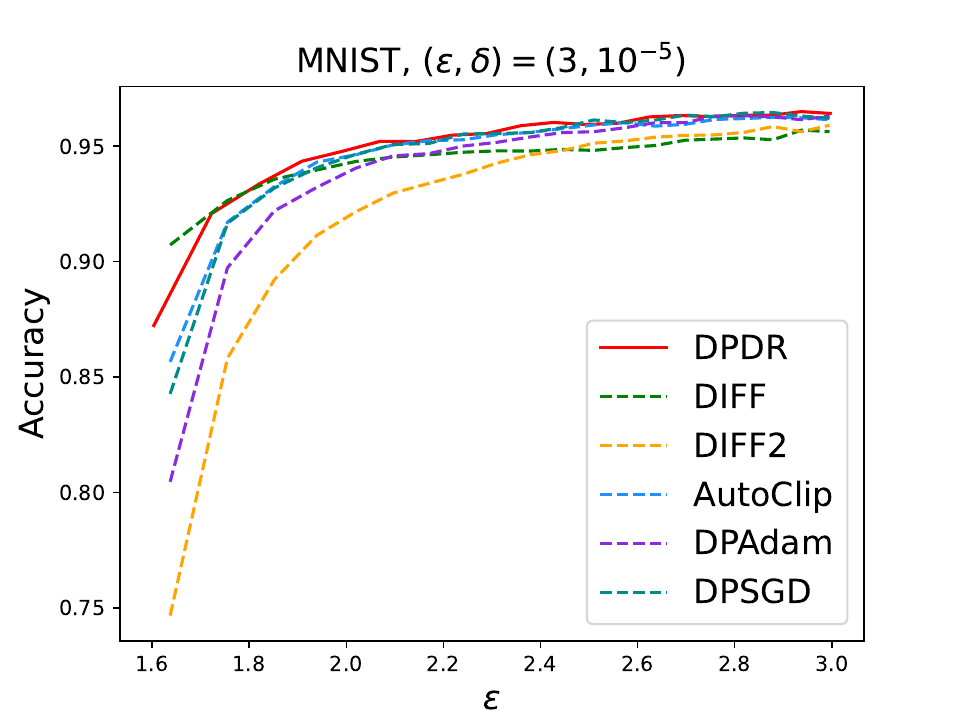}
  \includegraphics[width=0.32\textwidth, trim=10 5 30 10,clip]{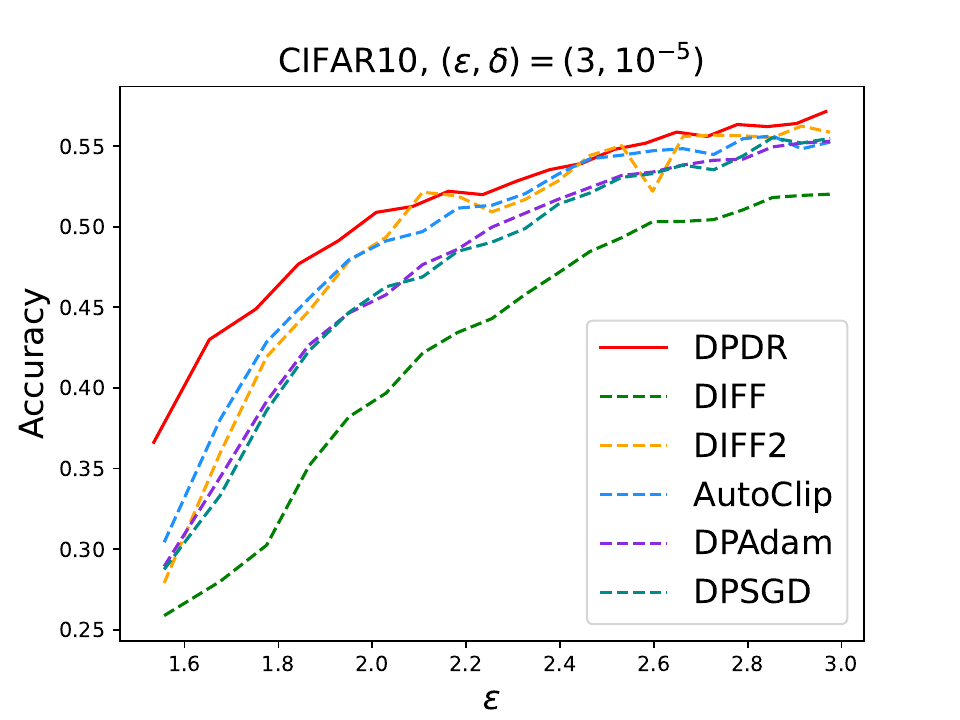}
  \includegraphics[width=0.32\textwidth, trim=10 5 30 10,clip]{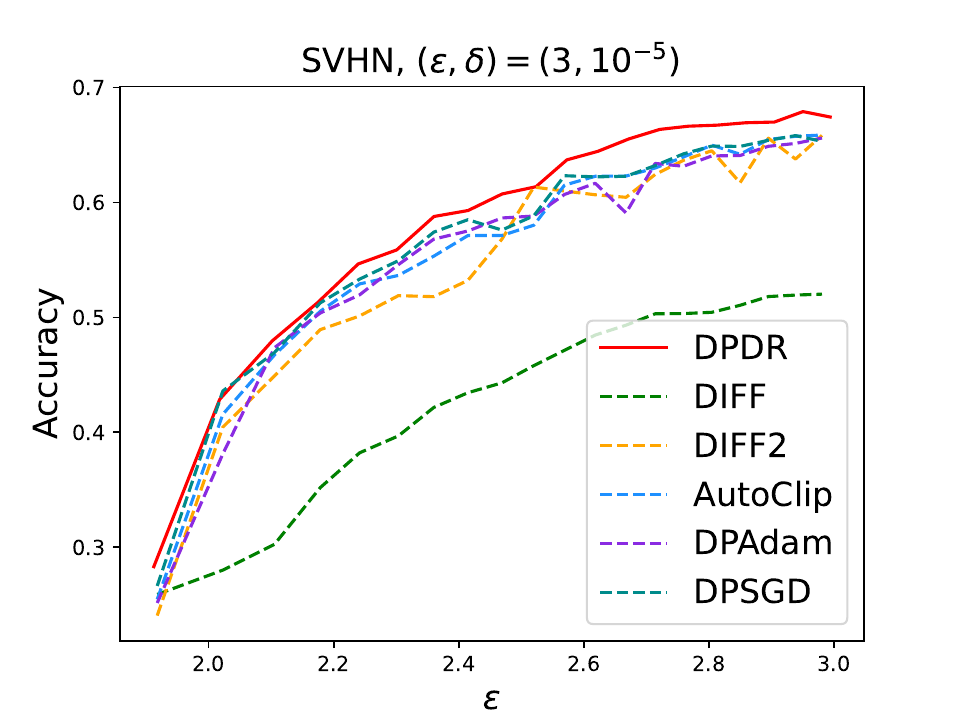}
  \caption{Convergence Evaluation  on CIFAR-10 with 5-layer CNN.}
\end{figure}

Fig. \ref{pics_conv} demonstrates the convergence rate of proposed method DPDR is higher than all other baselines. (1) Reaching the same accuracy, DPDR requires fewer steps, which indicates less privacy budgets. For instance, on CIFAR-10, DPDR consumes 2.59 $\epsilon$ for 55\% accuracy, while DP-SGD and its variants need 0.2 higher for same accuracy. (2) Though GDR only applies at the early steps, DPDR achieves consistently higher accuracy along all training steps. The promotion of DPDR comes from higher accuracy achieving by faster convergence rate of GDR at early steps, which provides a better starting point for following DP-SGD compared to the baseline model at the same step.

\vspace{-0.4cm}
\subsection{Effect of Hyperparameters}
\label{sec_exp_parameters}
\begin{figure}
  \centering
  \label{pics_param}
  \subfigure{
  \includegraphics[width=0.31\textwidth, trim=20 20 20 20,clip]{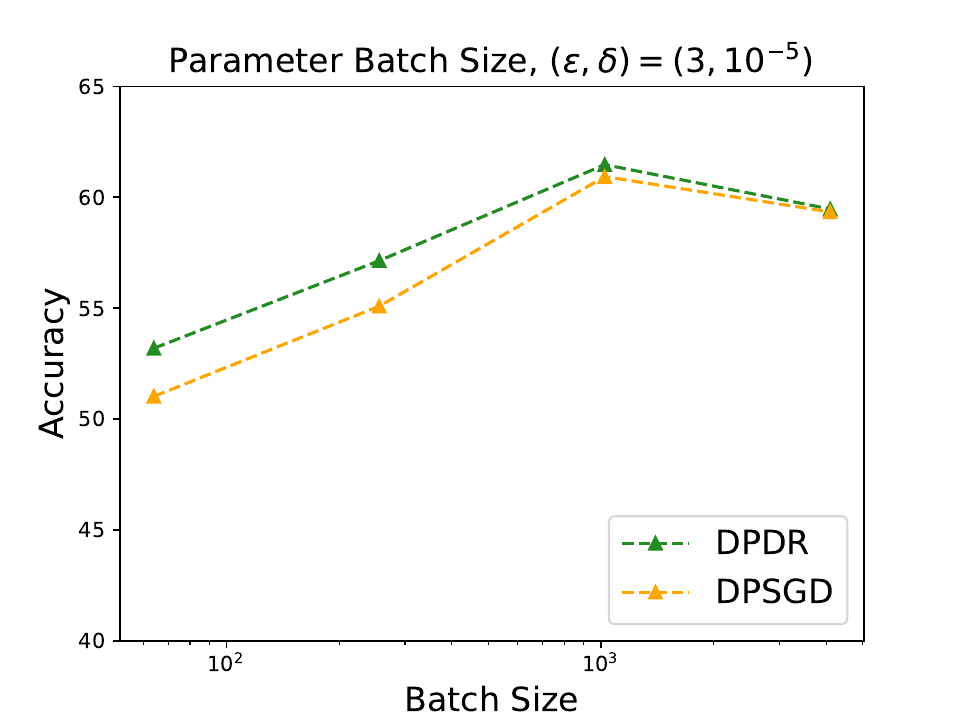}
  \label{pics_param_batchsize}
  }
  \subfigure{
  \includegraphics[width=0.31\textwidth, trim=20 20 20 20,clip]{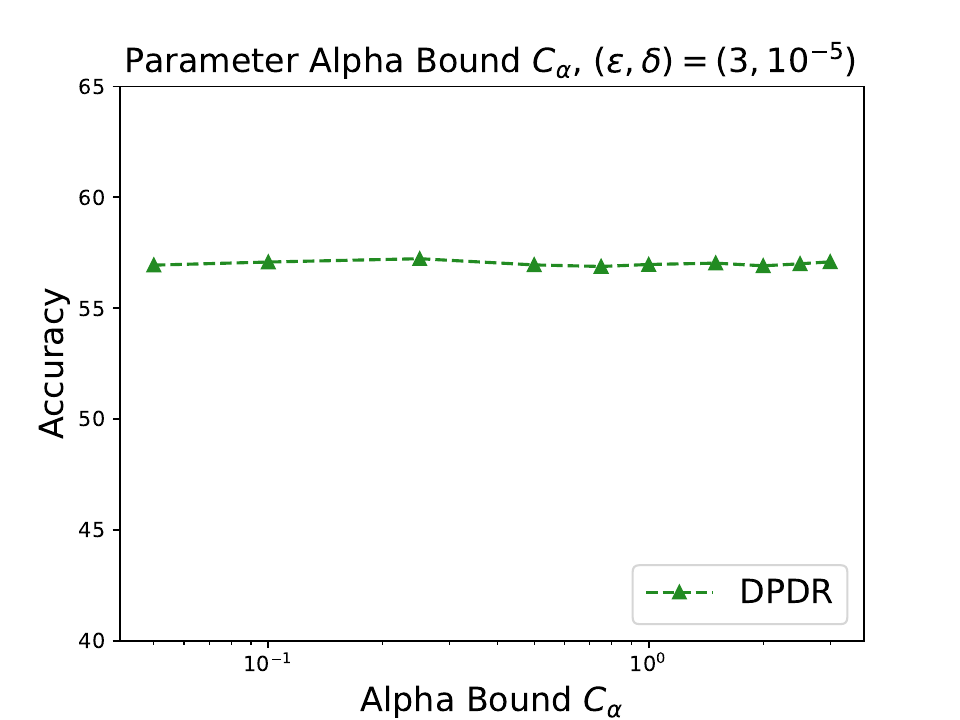}
  \label{pics_param_alphanorm}
  }
  \subfigure{
  \includegraphics[width=0.31\textwidth, trim=20 20 20 20,clip]{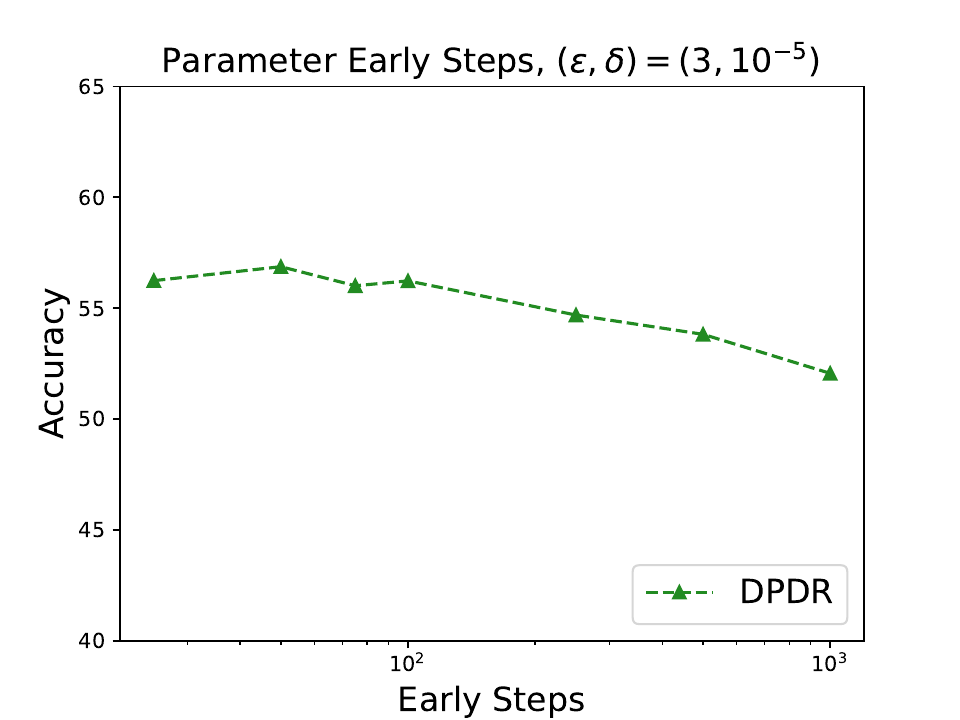}
  \label{pics_param_stepsdr}
  }
  \caption{Effect of Hyperparameters on CIFAR-10 with 5-layer CNN.}
\end{figure}

In Fig. \ref{pics_param}, we empirically study the effects of hyperparameters in DPDR, including batch size $B$, clipping bound of parallel coefficient $\alpha$, and the steps $s$ for early decomposition. As all datasets show similar properties over these parameters, we only show the impact for CIFAR-10 due to space. 

We evaluate the trade-off between batch size and accuracy in Fig. \ref{pics_param_batchsize}. (1) Both DP-SGD and proposed methods achieves the highest accuracy at $B=1024$, and gets lower as $B$ grows. The best accuracy achieved at a medium batch size is reasonable as larger batch size decreases and privacy amplification effect from subsampling ratio of batches, while smaller batch size introduce more sampling bias by stochastic gradient descent. (2) DPDR obtains higher accuracy across all batch sizes than DP-SGD, especially when batch size is small. It is reasonable as less samples introduce less incremental information, and common knowledge reused enjoys less perturbation due to strong privacy amplification effect. The result suggests the performance DPDR is more stable to smaller batch size, which is practical in reality considering the limitation of computational resources.

The impact of clipping bound of parallel coefficient $\alpha$ is demonstrated in Fig. \ref{pics_param_alphanorm}. The accuracy of DPDR is stable when $\alpha$ is over a wide range of $[0.05, 3]$. As smaller bound limits the effect of previous noisy gradients (common knowledge) and restrains the convergence rate, while reducing the DP noise amount injecting. While larger bound releashes the power of previous noisy gradients, but introduce more noises in training process. 

Fig. \ref{pics_param_stepsdr} demonstrates the effects of number of steps $s$ for gradient decomposition and reconstruction technique (GDR) at the beginning of DPDR. (1) The model achieves the highest accuracy at $s=50$, which is consistent with what we observed in Fig. \ref{pics_grad} (Right), orthogonal components are quite smaller than the entire gradients at the early training phases. (2) It is noticed that the accuracy decreases gradually when steps number gets larger. As common knowledge decreases along training process, the clipping and perturbing on parallel coefficient $\alpha$ introduce noises and bias in return for little information gain and still consuming privacy budgets. Hence at later training stages where gradient direction fluctuates, DP-SGD is more suitable for finding more elaborate optimal solution.

\section{Conclusion}
This work proposed DPDR, which focuses on enhancing the performance of DP-SGD by reducing the amount of noise injection in gradients. We achieved a higher information gain with a smaller amount of noise by introducing directional decomposition and reconstruction technique. The model accuracy is further enhanced by designing and leveraging a mixed strategy which makes the most of the privacy budget. Comprehensive experiments on real-world datasets and different models are conducted to confirm the effectiveness of DPDR on accuracy and convergence. In the future, we plan to extend DPDR to federated setting, and explore the effectiveness on advanced DP-SGD optimizers.

\bibliographystyle{plain}
\bibliography{main}

\appendix
\section{Appendix}
\subsection{Experiment Settings}
\label{appendix_exp}
\begin{table}
  \caption{Noise Multiplier for DPDR.}
  \label{tab_noise}
  \centering
  \begin{tabular}{*{5}{p{0.1\textwidth}<{\centering}}}
    \toprule
     $\epsilon$   & Dataset & $\sigma_\perp$ & $\sigma_\alpha$ & $\sigma_g$ \\
    \midrule
    $3$ & MNIST & 0.81 & 2.0 &  0.803  \\
     & CIFAR10  & 0.84 & 3.0  & 0.835  \\
     & SVHN      & 0.695 & 2.0 & 0.696   \\
    \midrule           
    $8$ & MNIST & 0.59 & 0.8 &  0.59  \\
     & CIFAR10  & 0.61 & 1.0  & 0.605  \\
     & SVHN      & 0.527 & 0.8 & 0.531   \\
    \bottomrule
  \end{tabular}
\end{table}
Table \ref{tab_noise} demonstrates the noise multipliers for DPDR and DPSGD. 

We conduct experiments on datasets MNIST\cite{lecun1998mnist}, CIFAR10\cite{krizhevsky2009learning} with model 4-layer CNN, 5-layer CNN which follows \cite{papernot2021tempered, tramer2020differentially}, and set ResNet18 on SVHN\cite{netzer2011reading} for evaluating DPDR on larger model. All of models adopts group normalization and cross-entropy loss function. 

We set all parameters as follows. The models are trained for 20 epochs under privacy budget $\epsilon=3$ and $8$ with fixed $\delta=10^{-5}$, batch size $B=256$. We tune hyperarameters clipping bound for each model and adopts result based on  best parameter combination. The tuning ranges are: $C_g \in [0.05, 1]$, $C_\perp \in [0.05, 1]$, $C_\alpha in {0.05, 1.5}$, learning rate $\gamma \in [0.1, 2]$. Specifically, for DIFF and DIFF2, the tuning range of clipping bound $C_d \in [0.001, 1]$. 

In DPDR, perturbing techique switches from GDR to DPSGD. The parameter selection on DPSGD phase is just the same as original DPSGD. We actually find out that the best clipping bounds and learning rates on original DPSGD and DPDR-DPSGD are the same.

\subsection{Proofs in Theoretical Analysis}
\subsubsection{Upper bound of sensitivity}
\label{appendix_sensitivity}
\paragraph{Sensitivity $\Delta f_\perp$} We analyze it from two angles. On the one hand, we generate the inequality based on triangle properties as follows:
\begin{align*}
	\Vert \Delta f_\perp \Vert
	= \Vert \nabla L(w_t) \cdot \sin(\nabla L(w_t), \tilde{\nabla L}(w_{t-1}) \Vert
	\leq \Vert \nabla L(w_t) \Vert
\end{align*}
 On the other hand, we could bound the noises with a more elaborate way:
\begin{align*}
      \Vert \Delta f_\perp \Vert
      &\leq (\Vert \nabla L(w_t) - \Pi_{\tilde{\nabla L}(x_{k-1})} (\nabla L(w_t)) \Vert^2 + \Vert \tilde{\nabla L}(x_{k-1}) - \Pi_{\tilde{\nabla L}(x_{k-1})} (\nabla L(w_t)) \Vert^2 )^{\frac{1}{2}} \\
      &= \Vert {\nabla L}(w_t) - \tilde{\nabla L}(w_{t-1}) \Vert 
      \leq \Vert \nabla L(w_t) - \nabla L(w_{t-1})\Vert  + \Vert \xi_{t-1} \Vert 
\end{align*}
Based on Chernoff inequality, we have
\begin{align*}
    \Pr(|\xi_{t-1} - \mathbb{E}[\xi_{t-1}]| < a) = \Pr(|\xi_{t-1}|< a) 
    \geq 1-\frac{\mathbb{D}[\xi_{t-1}]}{a^2} 
    \geq 1-\frac{(dC^2_\perp+mC^2_2)v|B|^2T\ln(1/\delta) }{N^2\epsilon_\perp^2 a^2}
\end{align*}

Considering the fact that real-world datasets usually satisfies that large dataset size, model size and small batch size, if $C_\perp=10^{-1}\Vert \nabla L(w_t) - \nabla L(w_{t-1}) \Vert$, with probability of at least $99\%$ we have $a<\Vert \nabla L(w_t) - \nabla L(w_{t-1}) \Vert$. Based on $a$ and the assumption that gradient satisfies $\rho$-Lipschitz condition that $\nabla L(w_t) - \nabla L(w_{t-1}) \leq \rho(w_t - w_{t-1})$,  $\Delta f_\perp$ is bounded with high probability that 
$\Vert \Delta f_\perp\Vert 
	\leq 2\Vert \nabla L(w_t) - \nabla L(w_{t-1})\Vert 
	\leq 2\rho \Vert w_t - w_{t-1} \Vert$.

Therefore, we achieve the upper bound for sensitivity of orthogonal components: 
\begin{align}
\label{eq_sensitivity_perp}
	\Vert \Delta f_\perp\Vert 
	\leq \min( \Vert \nabla L(w_t) \Vert,  2\rho \Vert w_t - w_{t-1} \Vert)
\end{align}

\paragraph{Sensitivity $\Delta f_\alpha$} The upper bound of sensitivity for parallel coefficient 
\begin{align*}
    \Vert \Delta f_\alpha \Vert = \frac{ \Vert <L(w_{t}), \dot{\tilde{L}}(w_{t-1})>\Vert}{\Vert \dot{{\tilde{L}}}(w_{t-1}) \Vert^2}
    = \Vert \cos \theta \cdot \nabla L(x_{t}) \Vert
    \leq \Vert \nabla L(w_{t}) \Vert
\end{align*}

\subsubsection{Theorem 2}
\label{appendix_conv}
\begin{proof}
According to the perturbation in DPDR, we have $\hat{g}_t = \frac{1}{B}\sum^B_{i=1} {\alpha}_{t} \cdot b + {g}_{\perp t} $ , $\tilde{g}_t =  \frac{1}{B} (\sum^B_{i=1}Clip(\alpha_t, C_\alpha)+N(0, C^2_\alpha\sigma^2_\alpha)) \cdot b + \frac{1}{B}(\sum^B_{i=1}Clip(g_{t\perp}, C_\perp)+N(0,C^2_\perp\sigma^2_\perp))$, where $b=\frac{\tilde{g}_{t-1}}{\Vert \tilde{g}_{t-1} \Vert}$. Under Assumption \ref{assumption_smooth} we have
\begin{align*}
    \mathcal{L}(w_{t+1}) &\leq \mathcal{L}(w_t) + <\nabla \mathcal{L}(w_t), w_{t+1}-w_t> + \frac{\rho}{2} \Vert w_{t+1} - w_t \Vert^2_2 \\
    &= \mathcal{L}(w_t) - \gamma<\nabla L(w_t), \tilde{g}_t> + \frac{\rho}{2}\gamma^2 \Vert \tilde{g}_t \Vert^2_2 
\end{align*}
Take expectation at both sides,
\begin{align}
\label{eq_conv}
\mathbb{E}[\mathcal{L}(w_{t})-\mathcal{L}(w_{t-1})]
\leq -\gamma \mathbb{E}[<\nabla \mathcal{L}(w_t), \tilde{g}_{t} >] +  \frac{\rho}{2}\mathbb{E}[\Vert \tilde{g}_{t} \Vert^2_2]
\end{align}
Now we analyze two terms on the right side of Eq.\eqref{eq_conv} separately.

For first term, 
\begin{align}
\label{eq_conv_1}
    \mathbb{E}[<\nabla L(w_t), \tilde{g_t}>] 
    = <\nabla L(w_{t-1}), \mathbb{E}[Clip(g_\perp, C_\perp)]> + <\nabla L(w_{t-1}), \mathbb{E}[Clip(\alpha, C_\alpha)\cdot b ]> 
\end{align}
The equation comes from the fact that DP noise $\mathbb{E}[N(0,a^2)]=0$ for arbitrary standard deviation $a$. We represent the sampling noise as $\xi_t=\xi_{\perp,t} + \xi_{\varparallel,t} = (g_{\perp,t}-\nabla\mathcal{L}_\perp(x_{t-1})) +(g_{\varparallel,t}-\nabla\mathcal{L}_{\varparallel}(x_{t-1}))  =g_t-\nabla\mathcal{L}(x_{t-1})$, and the probability of large sampling noise as $P_{\perp,t} = \Pr[\xi_{\perp,t}\in S_{\Vert \nabla\mathcal{L}_\perp(x_{t-1}) +\xi_{\perp,t}\Vert < C_\perp}]$. Then we have 
\begin{align}
\label{eq_conv2_perp}
    &<\nabla L(w_{t-1}), \mathbb{E}[Clip(g_\perp, C_\perp)]> \nonumber \\
    &= <\nabla L_\perp(w_{t-1}), \mathbb{E}[Clip(g_\perp, C_\perp)]> + <\nabla L_\varparallel(w_{t-1}), \mathbb{E}[Clip(g_\perp, C_\perp)]> \nonumber \\
    &= \mathbb{E}[\boldsymbol{1}_{\Vert \nabla\mathcal{L}_\perp(x_{t-1}) +\xi_{\perp,t}\Vert<C_\perp} (<\nabla\mathcal{L}_\perp(x_{t-1}), \nabla\mathcal{L}_\perp(x_{t-1})+\xi_{\perp,t}> + 0)]  \nonumber \\ 
    
    &+ C_\perp \mathbb{E}[\boldsymbol{1}_{\Vert \nabla\mathcal{L}_\perp(x_{t-1}) 
    +\xi_{\perp,t}\Vert \geq C_\perp} (<\nabla\mathcal{L}_\perp(x_{t-1}), \frac{\nabla\mathcal{L}_\perp(x_{t-1})+\xi_{\perp,t}}{\Vert \nabla\mathcal{L}_\perp(x_{t-1})+\xi_{\perp,t} \Vert}> 
    + 0)] \nonumber \\ 
    
    &\geq P_{\perp, t}\Vert \nabla\mathcal{L}_\perp(x_{t-1}) \Vert^2 - \Vert \nabla\mathcal{L}_\perp(x_{t-1}) \Vert \cdot \sqrt{(1-P_{\perp,t})\cdot \tau_{\perp,t}^2} + \mathbb{E}[\frac{C_\perp (1-P_{\perp,t})  \Vert \nabla\mathcal{L}_\perp(x_{t-1}) \Vert}{4} - \frac{15C_\perp\sqrt{(1-P_{\perp,t})}\tau_{\perp,t}}{4}]
\end{align}
The first equation comes from the fact that 
Similarly, for the parallel part we demote the probability of large sampling noise as $P_{\varparallel,t} = \Pr[\xi_{\varparallel,t}\in S_{\Vert \nabla\mathcal{L}_\perp(x_{t-1}) +\xi_{\varparallel,t}\Vert < C_\alpha}]$.
\begin{align}
\label{eq_conv3_paral}
    &<\nabla L(w_{t-1}), \mathbb{E}[Clip(\alpha, C_\alpha)\cdot b]> \nonumber \\
    &\geq \mathbb{E}[\boldsymbol{1}_{\Vert \nabla\mathcal{L}_\varparallel(x_{t-1}) +\xi_{\varparallel,t}\Vert<C_\alpha}<\nabla\mathcal{L}_\varparallel(x_{t-1}), \nabla\mathcal{L}_\varparallel(x_{t-1})+\xi_{\varparallel,t}>]  \nonumber \\ 
    
    &+ \mathbb{E}[\boldsymbol{1}_{\Vert \nabla\mathcal{L}_\varparallel(x_{t-1}) +\xi_{\varparallel,t}\Vert \geq C_\varparallel}<\nabla\mathcal{L}_\varparallel(x_{t-1}), \frac{\nabla\mathcal{L}_\varparallel(x_{t-1})+\xi_{\varparallel,t}>}{\Vert \nabla\mathcal{L}_\varparallel(x_{t-1})+\xi_{\varparallel,t} \Vert}] \nonumber \\ 
    &\geq P_{\varparallel, t}\Vert \nabla\mathcal{L}_\varparallel(x_{t-1}) \Vert^2 - \Vert \nabla\mathcal{L}_\varparallel(x_{t-1}) \Vert \cdot \sqrt{(1-P_{\varparallel,t})\cdot \tau_{\varparallel,t}^2} + \mathbb{E}[\frac{C_\alpha (1-P_{\varparallel,t})  \Vert \nabla\mathcal{L}_\varparallel(x_{t-1}) \Vert}{4} - \frac{15C_\alpha\sqrt{1-P_{\varparallel,t}}\tau_{\varparallel,t}}{4}]
\end{align}
the first inequatility results from when clipping happens, $\frac{\alpha}{\Vert \alpha \Vert} \cdot b \geq \frac{\alpha \cdot b}{\Vert \alpha \cdot b \Vert} \geq \frac{ \nabla\mathcal{L}_\varparallel(x_{t-1}) +\xi_{\varparallel,t}}{\Vert \nabla\mathcal{L}_\varparallel(x_{t-1}) +\xi_{\varparallel,t} \Vert} $
Take Eq. \eqref{eq_conv2_perp} and \eqref{eq_conv3_paral} back into Eq. \eqref{eq_conv_1}, we obtain the first term of Eq. \eqref{eq_conv}

For second term,
\begin{align*}
    &\mathbb{E}[\Vert \tilde{g}_t \Vert^2_2]
    = \mathbb{E}[\Vert (Clip(g_\perp, C_\perp)] + \mathbb{E}[\Vert \eta_\perp\Vert^2] + \mathbb{E}[ \Vert Clip(\alpha, C_\alpha)\cdot b \Vert^2] + \mathbb{E}[\Vert \eta_\alpha\cdot b \Vert^2] \\
    &\leq \frac{1}{n^2q^2} (C_\perp (n(n-1)q^2+qn) + dC^2_\perp\sigma^2_\perp + C_\alpha (n(n-1)q^2+qn) + mC^2_\alpha\sigma^2_\alpha)\\
\end{align*}
Overall,
\begin{align}
\label{eq_conv_end1}
\mathbb{E}[<\nabla \mathcal{L}(w_t), \tilde{g}_{t} >]
\leq \frac{1}{\gamma}\mathbb{E}[\mathcal{L}(w_{t})-\mathcal{L}(w_{t-1})] + \frac{\rho\gamma}{2}\mathbb{E}[\Vert \tilde{g}_{t} \Vert^2
\end{align}

\begin{align*}
\label{eq_conv_end2}
&P_{\perp, t}\Vert \nabla\mathcal{L}(x_{t-1}) \Vert^2 + \Vert \nabla\mathcal{L}_\perp(x_{t-1}) \Vert \cdot(\frac{C_\perp (1-P_{\perp,t})}{4}- \sqrt{1-P_{\perp,t}} \tau_{\perp,t}) \\
&+ \Vert \nabla\mathcal{L}_\varparallel(x_{t-1}) \Vert \cdot(\frac{C_\varparallel (1-P_{\varparallel,t})}{4}- \sqrt{1-P_{\varparallel,t}} \tau_{\varparallel,t}) \\
& \leq \frac{1}{\gamma}\mathbb{E}[\mathcal{L}(w_{t})-\mathcal{L}(w_{t-1})] + \frac{\rho\gamma}{2}\mathbb{E}[\Vert \tilde{g}_{t} \Vert^2 + \mathbb{E}[\frac{15C_\perp\sqrt{(1-P_{\perp,t})}\tau_{\perp,t}}{4} + \frac{15C_\alpha\sqrt{(1-P_{\alpha,t})}\tau_{\alpha,t}}{4}]
\end{align*}

Considering T steps,
\begin{align*}
&\mathbb{E}[P_{\perp, t}\Vert \nabla\mathcal{L}(x_{t-1}) \Vert^2 + \Vert \nabla\mathcal{L}_\perp(x_{t-1}) \Vert \cdot(\frac{C_\perp (1-P_{\perp,t})}{4}- \sqrt{1-P_{\perp,t}} \tau_{\perp,t}) \\
&+ \Vert \nabla\mathcal{L}_\varparallel(x_{t-1}) \Vert \cdot(\frac{C_\varparallel (1-P_{\varparallel,t})}{4}- \sqrt{1-P_{\varparallel,t}} \tau_{\varparallel,t})] \\
& \leq \frac{1}{\gamma T}\mathcal{L}(w_{0}) + \frac{\rho\gamma'}{2}(2C_\perp^2+dC_\perp^2\sigma_\perp^2+2C_\alpha^2+mC_\alpha^2\sigma_\alpha^2) \\
&+ \frac{15}{4T}\sum_{t=1}^T\mathbb{E}[C_\perp\tau_{\perp,t}\sqrt{(1-P_{\perp,t})} + C_\alpha\tau_{\alpha,t}\sqrt{(1-P_{\alpha,t})}] \\
& = \leq \frac{1}{\gamma T}\mathcal{L}(w_{0}) + \mathcal{O}(\rho\gamma d C^2_\perp \sigma^2_\perp + m C^2_\alpha \sigma^2_\alpha) 
+ \frac{15}{4T}\sum_{t=1}^T\mathbb{E}[C_\perp\tau_{\perp,t}\sqrt{(1-P_{\perp,t})} + C_\alpha\tau_{\alpha,t}\sqrt{(1-P_{\alpha,t})}]
\end{align*}

\end{proof}

\end{document}